\pdfoutput=1

\documentclass[11pt]{article}

\usepackage[final]{acl}
\usepackage{multirow}
\usepackage{times}
\usepackage{latexsym}
\usepackage{tcolorbox}
\usepackage{subcaption}
\usepackage[T1]
{fontenc}

\usepackage[utf8]{inputenc}

\usepackage{microtype}
\usepackage{colortbl}
\usepackage{marvosym}

\usepackage{inconsolata}

\usepackage{graphicx}
\usepackage{xspace}
\usepackage{amsmath}
\usepackage{enumitem}
\usepackage{booktabs}

\newcommand{\Ours}[0]{Amber\xspace}

\title{Towards Adaptive Memory-Based Optimization for Enhanced Retrieval-Augmented Generation}

\author{%
   Qitao Qin\textsuperscript{1*}, Yucong Luo\textsuperscript{1}\thanks{These authors contributed equally to this work.},Yihang Lu\textsuperscript{1}, Zhibo Chu\textsuperscript{1},
   Xiaoman Liu\textsuperscript{1},
   Xianwei Meng\textsuperscript{2}\textsuperscript{\Letter}
   \\
   \textsuperscript{1}University of Science and Technology of China \\  \textsuperscript{2}	Hefei Institutes of Physical Science, Chinese Academy of Sciences \\
  \texttt{\{qqt,prime666,lyhsa22,zb.chu,liuxiaoman\}@mail.ustc.edu.cn} \\ \texttt{mengxw@iim.ac.cn}\\
  }

\begin{document}
\maketitle

\begin{abstract}
Retrieval-Augmented Generation (RAG), by integrating non-parametric knowledge from external knowledge bases into models, has emerged as a promising approach to enhancing response accuracy while mitigating factual errors and hallucinations. This method has been widely applied in tasks such as Question Answering (QA). 
However, existing RAG methods struggle with open-domain QA tasks because they perform independent retrieval operations and directly incorporate the retrieved information into generation without maintaining a summarizing memory or using adaptive retrieval strategies, leading to noise from redundant information and insufficient information integration.
To address these challenges, we propose \textbf{A}daptive \textbf{m}emory-\textbf{b}ased optimization for \textbf{e}nhanced \textbf{R}AG (\textbf{\Ours}) for open-domain QA tasks, which comprises an Agent-based Memory Updater, an Adaptive Information Collector, and a Multi-granular Content Filter, working together within an iterative memory updating paradigm. Specifically, \textbf{\Ours} integrates and optimizes the language model's memory through a multi-agent collaborative approach, ensuring comprehensive knowledge integration from previous retrieval steps. It dynamically adjusts retrieval queries and decides when to stop retrieval based on the accumulated knowledge, enhancing retrieval efficiency and effectiveness. Additionally, it reduces noise by filtering irrelevant content at multiple levels, retaining essential information to improve overall model performance. We conduct extensive experiments on several open-domain QA datasets, and the results demonstrate the superiority and effectiveness of our method and its components. 
The source code is available \footnote{https://anonymous.4open.science/r/Amber-B203/}.
\end{abstract}

\section{Introduction}

In recent years, Large Language Models (LLMs) ~\cite{brown2020language,achiam2023gpt,touvron2023llama,anil2023palm} have demonstrated exceptional performance across various tasks, including question answering (QA)~\cite{yang2018hotpotqa,kwiatkowski2019natural}, owing to their ability to capture diverse knowledge through billions of parameters. However, even the most advanced LLMs often suffer from hallucinations~\cite{chen2023chatgpt} and factual inaccuracies due to their reliance on parametric memory. Additionally, it is impractical for these models to memorize all of the ever-evolving knowledge.
To address these challenges, retrieval-augmented generation (RAG)~\cite{borgeaud2022improving,izacard2023atlas,shi2023replug} have garnered increasing attention. These models retrieve passages relevant to the query from external corpora and incorporate them as context to the LLMs, enabling the generation of more reliable answers. By integrating retrieved information, retrieval-augmented LLMs maintain both the accuracy and timeliness of their knowledge.
Early studies on RAG primarily focused on single-hop queries~\cite{lazaridou2022internet,ram2023context}, where answers can typically be found within a single document. However, these methods often fall short when handling complex QA tasks, such as long-form QA and multi-hop QA, which require aggregating information from multiple sources. Unlike single-hop QA, these queries necessitate connecting and synthesizing information across multiple documents and cannot be solved by a single retrieval-and-response step. For instance, the query ``\textit{Is Microsoft Office 2019 available in a greater number of languages than Microsoft Office 2013?}'' requires three reasoning steps: first, retrieving information about the languages supported by ``\textit{Office 2019}''; second, retrieving similar information for ``\textit{Office 2013}''; and finally, comparing the two sets of data to produce an answer.

To address this issue, Adaptive RAG has been proposed. It adaptively selects appropriate retrieval questions and timing based on the difficulty of the user query to flexibly capture more valuable knowledge for answering open-domain QA tasks, achieving a balance between effectiveness and efficiency. However, these methods still have several problems. 
\textbf{\textit{First}}, each retrieval operates independently and lacks a summarizing memory of previous retrieval fragments, which may cause the outputs to reflect only limited knowledge from specific retrieval steps while neglecting the integration and interaction of retrieved information from different steps.
\textbf{\textit{Second}}, when the LLM uses these retrieved fragments for reasoning, it does not actively evaluate the validity of the information. Consequently, without the ability to determine when to proactively stop retrieval based on known information or update the queries that need to be retrieved, it may lead to inefficiencies or the retrieval of irrelevant information.
\textbf{\textit{Third}}, the effective parts within the retrieved text segments are very few, and excessive redundant information introduces noise, which can obscure important details and negatively impact the model's performance.

To this end, we propose \textbf{A}daptive \textbf{m}emory-\textbf{b}ased optimization for \textbf{e}nhanced \textbf{R}AG (\textbf{\Ours}). \Ours comprises three core components: Agent-based Memory Updater (AMU), Adaptive Information Collector (AIC), and Multi-granular Content Filter (MCF). These components work in unison to automatically integrate and update retrieved information as the LLM's memory, dynamically adjust the queries based on known information, and employ multi-granular content filtering during retrieval to retain useful information and reduce noise, thereby achieving outstanding performance. 
\textbf{\textit{Firstly}}, to address the issue in which each retrieval operates independently and lacks a summarizing memory of previous retrieval fragments, AMU employs a multi-agent collaborative approach. By coordinating various agents, AMU optimizes the LLM's current memory. This process ensures that the knowledge structure is continuously refined and enriched, effectively integrating all valuable information from previous retrieval steps.
\textbf{\textit{Secondly}}, AIC utilizes the real-time memory generated by AMU to update the queries that need to be retrieved and decides when to stop retrieval. By automatically adjusting the retrieval process based on the accumulated knowledge, AIC ensures that subsequent retrievals are more targeted and efficient, effectively addressing the challenge of insufficient knowledge accumulation and avoiding unnecessary retrievals.
\textbf{\textit{Lastly}}, we fine-tune the LLM to function as MCF to reduce noise during retrieval. MCF includes two levels of filtering capabilities. Firstly, it assesses the validity of the entire retrieved text segment and the query, determining whether the information is relevant and useful. Secondly, from the valid retrieved segments, it filters out irrelevant content and retains essential information. This approach effectively reduces redundant information and highlights crucial details, thereby enhancing the overall performance of the model.

In summary, our contributions are as follows.
\begin{itemize}[left=0.em, itemsep=-5pt, topsep=5pt]
\item We propose the Agent-based Memory Updater, which uses a multi-agent approach to integrate information and form memory from previous retrievals, optimizing the LLM's memory.

\item We develop the Adaptive Information Collector, which updates retrieval queries and decides when to stop retrieval, making the process more targeted and efficient.

\item We introduce the Multi-granular Content Filter to reduce noise by filtering irrelevant content at multiple levels, enhancing model performance.

\item Extensive experiments validate the effectiveness of \Ours, showing significant improvements over existing methods in open-domain QA.
\end{itemize}

\begin{figure*}[h] 
    \centering
    \includegraphics[width=1\textwidth]{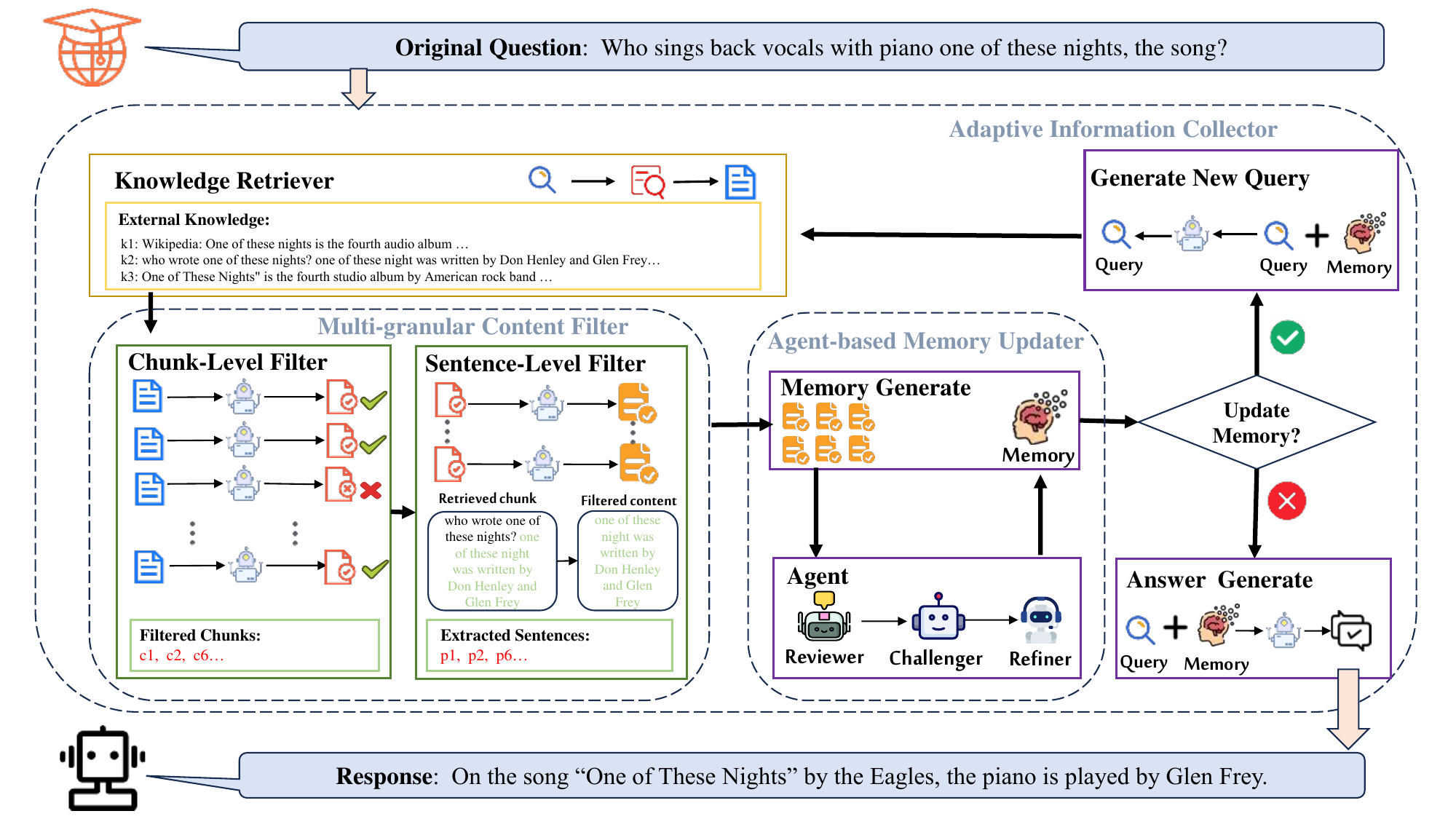} 
    \vspace{-0.3in}
    \caption{
\textbf{Illustration of the \Ours framework.} \Ours is an adaptive Retrieval-Augmented Generation (RAG) approach incorporating three key components: the Adaptive Information Collector (AIC), the Multi-Granular Content Filter (MCF), and the Agent-Based Memory Updater (AMU). The MCF filters chunks irrelevant to the query and extracts the most useful sentences. Subsequently, the AMU updates the generated memory notes. Finally, the AIC evaluates the quality of the memory and determines whether further iterations are necessary.} 
    \label{fig:1_framework} 
    \vspace{-0.2in}
\end{figure*}
\section{Related Work}
\paragraph{Open-domain 
 QA}
 Open-domain Question Answering (OpenQA)~\cite{voorhees1999trec} seeks to provide answers to questions expressed in natural language that are not restricted to a specify domain. Modern methods for Open QA tasks typically adopt the Retriever-and-Reader framework~\cite{chen2017reading,wang2024self}. With the advancement of open-domain QA, multi-hop QA ~\cite{joshi2017triviaqa,yang2018hotpotqa} and long-form ~\cite{stelmakh2022asqa,lyu2024crud} have gradually emerged. This more complex QA further necessitates the system to extensively collect and contextualize from the multiple documents (typically through iterative processes) in order to address more intricate queries. In particular, ~\cite{khot2022decomposed,khattab2022demonstrate} suggested breaking down multi-hop queries into simpler single-hop queries, iteratively utilizing LLMs and the retriever to address these sub-queries, and then combining their results to construct a comprehensive answer. Unlike the decomposition-based method, other recent studies, such as ~\cite{yao2022react} and ~\cite{trivedi2022interleaving}, explored a technique that creates a logical sequence of reasoning steps with document retrieval. Additionally, ~\cite{jiang2023active} proposed a method that involves iteratively fetching new documents when the tokens in the generated sentences exhibit low confidence, and ~\cite{jeong2024adaptive} proposed a retrieval strategy based on the complexity of the questions. However, The methods mentioned above neglect both the quality of retrieved documents and the generation of memory. Therefore, it is essential to propose a method aimed at enhanceing the quality of both retrieval and memory generation.
\paragraph{Retrieval-Augmented Generation.}
RAG has become essential for enhancing the response quality of large language models (LLMs). Early strategies~\cite{izacard2023atlas} relied on single-time retrieval, inputting the retrieved passages directly into LLMs to generate answers. However, these methods often fell short in complex tasks like multi-hop and long-form question answering, failing to capture sufficient information. To address these limitations, multi-time retrieval ~\cite{trivedi2022interleaving,borgeaud2022improving} was explored, though it risked incorporating irrelevant data, leading to poor-quality responses. This led to the development of Adaptive RAG (ARAG), which dynamically adjusts retrieval strategies based on real-time feedback, determining optimal retrieval times and content. Key innovations include Flare~\cite{jiang2023active}, which triggers new retrievals for low-confidence tokens, and Self-RAG~\cite{asai2023self}, which uses self-reflective markers to assess content quality. These adaptive approaches enhance retrieval relevancy and accuracy, while agent-based models like ReAct~\cite{yao2022react} further augment RAG's flexibility and intelligence. Nevertheless, we argue that the above methods overlook the quality issues in both retrieval and agent generation, resulting in inaccurate answers.

\section{Methods}
In this section, we define the task and present an overview of our proposed method, \Ours (illustrated in Figure 1). Following this, we provide a detailed explanation of each individual component.

\subsection{Problem Formulation}
RAG aims to enhance the generation quality of LLMs by integrating relevant information from an external document corpus \(D = \{d_1, d_2, \dots, d_n\}\). Given a user input \(x\) or a query \(q\), the core of RAG involves using a retriever \(R\) to identify and select a subset of pertinent documents from \(D\). The LLM then leverages both the original input and these retrieved documents to produce an improved output. Generally, this process seeks to achieve an output \(y\) based on the input and retrieved context.

\subsection{\Ours Overview}
Figure \ref{fig:1_framework} presents the architecture of \Ours, an adaptive memory updating iterative RAG method. It consists of three main components: an Agent-based Memory Updater, an Adaptive Information Collector, and a Multi-granular Content Filter.

Given a query \(q\), we initialize the LLM's memory \(M_0\) as an empty set. Acting as the primary scheduler, the \textbf{Adaptive Information Collector} initiates the iterative loop. At each iteration \(t\), we retrieve the top \(k\) text chunks \(C_t = \{ c_1^t, c_2^t, \dots, c_k^t \}\) from the corpus \(D\) based on \(q\). The \textbf{Multi-granular Content Filter} then assesses the relevance of each chunk \(c_i^t\) to \(q\) and filters out irrelevant content, retaining the most pertinent sub-paragraphs to form the refined document set \(P_t\).
Subsequently, the \textbf{Agent-based Memory Updater} employs three agents—the Reviewer, Challenger, and Refiner—to collaboratively integrate the new references \(P_t\) with the previous memory \(M_{t-1}\), producing the updated memory \(M_t\). These agents ensure that the memory is optimized for answering \(q\).
The adaptive information collector then evaluates whether the LLM can adequately respond to \(q\) using the current memory \(M_t\). If not, it generates a new retrieval query \(q_{t+1}\) based on \(M_t\) and \(q\) and proceeds to the next iteration. If the LLM can answer \(q\) satisfactorily, the adaptive information collector terminates the loop.
After the iterative process concludes, we use the final memory \(M_T\) as context to generate the answer \(\alpha \in A\) through the LLM's zero-shot in-context learning (ICL). 

\subsection{Agent-based Memory Updater (AMR)}

In real-world scenarios, user queries vary significantly in complexity, necessitating the formulation of tailored strategies for each query. Enhanced LLM based on memory provide an effective solution to this challenge. Memory, which represents the information known to the LLM during retrieval, enables the model to determine retrieval strategies effectively. Among these, memory updating is a critical component. It requires the LLM to leverage historical and newly retrieved information to regenerate the latest memory aligned with the query.

Inspired by the concept of multi-agent collaboration, we propose an \textbf{Agent-based Memory Updater} framework, which employs a cooperative, multi-agent approach to memory updating. Specifically, AMR consists of three independent agents: the Reviewer, the Challenger, and the Refiner. Through iterative dialogue, these agents reflect upon and optimize the memory. The inputs to AMR include the current memory \( m_t \), the retrieved passages \( p_t \) for the current query \( q_t \), and the original user query \( q \). Based on these inputs, the LLM initially generates an updated memory \( m_{t+1} \).

\textbf{Reviewer.} As the primary evaluator in the AMR framework, the Reviewer examines the proposed memory update \( m_{t+1} \) using the current memory \( m_t \), retrieved passages \( p_t \), and user query \( q \). The Reviewer assesses the correctness and relevance of \( m_{t+1} \) to the user's intent, identifying strengths and weaknesses. By sharing evaluations with the Challenger and Refiner, the Reviewer facilitates collaborative refinement and coordinates strategies to ensure alignment with collective goals. This evaluation process ensures memory updates are rigorously reviewed, improving the LLM's retained information.

\textbf{Challenger.} Acting as the critical analyst, the Challenger builds upon the Reviewer's assessment by examining \( m_{t+1} \), identifying potential flaws and overlooked constraints. Through interaction with the Reviewer and Refiner, the Challenger scrutinizes the validity of the memory update, raising probing questions about conflicts with existing knowledge or unmet user requirements. These interactions enable collective strategy adaptation, ensuring \( m_{t+1} \) is robust and well-aligned with both the user query and knowledge base.

\textbf{Refiner.} As the agent responsible for implementing improvements, the Refiner synthesizes feedback from the Reviewer and Challenger to refine \( m_{t+1} \). It translates critiques into concrete modifications, focusing on enhancing accuracy, clarity, and adherence to user query. The Refiner resolves issues identified by other agents, producing a revised \( m_{t+1} \) that better satisfies objectives. Through collaboration with the Reviewer and Challenger, the Refiner streamlines the feedback loop and maintains modification records, contributing to an effective refinement cycle.

Through the complementary collaboration of the Reviewer, Challenger, and Refiner within AMR, the proposed method effectively leverages the strengths of each agent. This triadic interaction ensures that memory updates undergo rigorous evaluation, critical examination, and precise refinement. As a result, the updated memory \( m_{t+1} \) becomes increasingly accurate, relevant, and aligned with the user’s query across multiple iterative cycles.

\subsection{Adaptive Information Collector (AIC)}
We propose the \textbf{Adaptive Information Collector} as the primary scheduler to control the entire RAG workflow. The role of AIC is to evaluate whether the information currently available in the memory, generated by the AMU, is sufficient to answer the user query \( q \).

Specifically, each iteration of AIC follows three key steps. The process initializes with the user query \( q_0 = q \) and an empty memory \( m_0 = \emptyset \). Firstly, AIC begins by retrieving the top \( k \) text chunks \( C_t = \{ c_1^t, c_2^t, \dots, c_k^t \} \) from the corpus \( D \) based on the current query \( q_t \) using a retrieval mechanism. Next, the query \( q \), the retrieved text chunks \( C_t \), and the current memory \( m_t \) are input into the Agent-based Memory Updater (AMU), which generates an updated memory \(m_{t+1} = \text{AMU}(q_t, C_t, m_t)\). Lastly, AIC then evaluates whether the updated memory \( m_{t+1} \) contains sufficient information to fully answer the query \( q \). If the memory is deemed sufficient, the iterative process terminates, and the latest memory \( m_T \), along with the query \( q \), are inputted into the LLM using in-context learning to produce the final answer \( a \). However, if the updated memory \( m_{t+1} \) is insufficient, AIC generates a refined query \(q_{t+1} = \text{AIC}(q, q_t, m_{t+1}).\) to target the missing information and proceeds to the next iteration. This iterative approach ensures that the final memory \( m_T \) is comprehensive and well-aligned with the user’s informational needs.

This iterative design allows AIC to dynamically refine queries and memory updates, ensuring that the final memory \( m_T \) contains the necessary information to answer the user’s query comprehensively.

\subsection{Multi-granular Content Filter (MCF)}

The Adaptive Information Collector, despite leveraging the filtering capabilities of the Agent-based Memory Updater and employing adaptive retrieval to refine the query \( q_t \), often retrieves the top \( k \) text chunks \( C_t = \{ c_1^t, c_2^t, \dots, c_k^t \} \) that still include irrelevant information. These irrelevant parts can be categorized into two levels: chunk-level irrelevance, where an entire chunk \( c_i \) may be unrelated to the query \( q \), and sentence-level irrelevance, where even within a relevant chunk \( c_i \), only a subset of the sentences may be pertinent to the query \( q \), while the remainder constitutes noise.

Based on these insights, we used STRINC, CXMI metrics, and GPT-4 (detail see in appendix \ref{appendix:filter}) to generate a multi-granular content filter dataset and subsequently fine-tuned a LLM using multi-task learning to create the \textbf{Multi-granular Content Filter}, denoted as \( F_c \). This content filter operates hierarchically, applying two levels of filtering to each chunk \( c_i \).

At the first level, a chunk-level filtering prompt, formulated as \( f_c(prompt_{chunk}, q, p_i) \), determines whether a chunk is relevant to \( q \). If \( f_c \) returns False, the chunk is directly discarded; otherwise, it progresses to the second level. The chunk-level filter is defined as:

\begin{small}
\begin{equation}
f_c({\small prompt_{chunk}}, q, c_i) = 
\begin{cases} 
\text{\small True}, & \text{\small if } c_i \text{\small \  relevant to } q \\
\text{\small False}, & \text{\small if } c_i \text{\small \ not relevant to } q
\end{cases}
\end{equation}
\end{small}

\noindent At the second level, a sentence-level evaluation is performed for chunks that pass the initial filter, where \( p_i = f_c(prompt_{sentence}, q, c_i) \) assesses each sentence within the chunk to retain the relevant sentences. The output of this stage is a refined set of relevant sentences \( P_t = \{ p_1, p_2, \dots, p_m \} \), where \( p_i \) are the relevant sentences.

This hierarchical filtering approach significantly reduces noise in the retrieved information by isolating only the relevant content at both chunk and sentence levels. The MCF, ensures that AIC operates with higher precision, improving the quality and relevance of the memory \( m_t \) in each iteration and, consequently, the overall performance.



\section{Experimental Setup}
In this section, we present the datasets, models, metrics, and implementation details.
More experiment setup can see appendix \ref{appendix:filter} and \ref{appendix:retriever}.
\subsection{Datasets and Evaluation Metrics}

To simulate a realistic scenario, where different queries have varying complexities, we use both the single-hop, multi-hop and long-form QA datasets simultaneously, in the unified experimental setting.
\paragraph{Single-hop QA}
\begin{table*}[h]
\centering
\caption{\textbf{Performance comparison of \Ours with baseline models.} The bold and underlined values indicate the best and second-best results across all models. Overall, \Ours consistently achieves superior performance across all datasets, demonstrating its effectiveness in answering questions.}
\vspace{-0.1in}
\label{tab:2_main_results}
\scalebox{0.64}{
\begin{tabular}{@{}llcccccccccccc@{}}
\toprule
\multirow{3}{*}{} & \multirow{3}{*}{Methods} & \multicolumn{6}{c}{single-hop QA} & \multicolumn{4}{c}{multi-hop QA} & \multicolumn{2}{c}{Long-form QA} \\
\cmidrule(lr){3-8} \cmidrule(lr){9-12} \cmidrule(lr){13-14}
 &  & \multicolumn{2}{c}{SQuAD} & \multicolumn{2}{c}{Natural Questions} & \multicolumn{2}{c}{TriviaQA} & \multicolumn{2}{c}{2WikiMQA} & \multicolumn{2}{c}{HotpotQA} & \multicolumn{2}{c}{ASQA}\\
 
 &  & acc & f1 & acc & f1 & acc & f1 & acc & f1 & acc & f1 & str-em & str-hit \\
\midrule
\multirow{1}{*}{No Retrieval} 
 & NoR   & 12.6  & 18.41 & 24.0 & 27.49 & 49.8 & 52.69 & 28.4 & 35.6 & 19.8  & 25.17 & 35.5 & 8.9 \\
\midrule
\multirow{6}{*}{Single-time RAG} 
 & Vanilla (Qwen2-7b) & 32.2 & 27.7 & 36.2 & 24.62 & 60.6 & 49.63 & 36.2 & 39.0 & 37.8 & 37.2 & 43.5 & 18.5 \\
 & Vanilla (Llama3-8b)& 30.4 & 36.08 & 33.2 & 38.99 & 58.2 & 60.28 & 22.2 & 26.2 & 34.2 & 42.2 & 38.7 & 13.7 \\
  & Vanilla (GPT-3.5)& 34.4 & 37.88 & 35.9 & 38.43 & 63.8 & 63.49 & 35.4  & 38.2 & 38.6  & 44.36 & 47.77 & 21.62\\
 & Self-Refine & 32.1 & 33.04 & 35.8 & 35.17 & 61.2 & 58.91 & 35.9 & 38.6 & 38.2 & 43.8 & 42.1 & 16.6 \\
 & Self-Rerank & 31.1 & 35.19 & 34.3 & 39.05 & 60.7 & 59.84 & 34.8 & 32.1 & 35.6 & 42.2 & 35.0 & 11.4 \\
 & Chain-of-note  & 31.8  & 33.94 & 35.2 & 37.66 & 61.0 & 58.33 & 35.1 & 39.7 & 36.8  & 45.0 & 40.3 & 15.6 \\
\midrule
\multirow{6}{*}{Adaptive RAG} 
  & ReAct  &  33.6& 34.85 & 35.4 & 38.37 & 60.9 & 59.83 & 34.6 & 37.3 & 37.5 & 46.9 & 32.9 & 8.3 \\
  & Self-RAG  & 32.7  & 33.84 & 37.9 & 39.17 & 60.3 & 58.94  &  29.8 & 30.8 & 35.3  & 44.4 & 40.9 & 16.5 \\
  & FLARE & 32.9 & 35.81 & 36.4 & 38.94 & 61.1 & 57.75 & 38.2 & 42.8 &  37.2 & 47.8 & 34.9 & 9.5 \\
  & Adaptive-RAG & 33.0 & 38.3 & 44.6 & 47.3 & 58.2 &60.7  & 46.4 & \underline{49.75} & 44.4 & 52.56 & 42.1 & 15.8 \\
  & Adaptive-Note & 29.0 & 33.61 & 40.0 & 45.38 & 59.6 & 59.72 & 39.4 & 39.1 & 39.0 & 46.6 & 43.7 & 17.7 \\
\midrule
\multirow{3}{*}{Ours} 
 & \Ours (Qwen2-7b)  & \textbf{36.8} & 38.43 & \textbf{47.8} & 49.84 & \underline{65.8} & 62.77 & \textbf{56.0} & \textbf{52.73} & \textbf{52.6} & 51.13 & \underline{49.7}  & \underline{25.2}\\
 & \Ours (Llama3-8b)  & 34.6 & \textbf{39.37} & 44.2 & \underline{50.49} & 63.6 & \underline{62.79} & 43.8 & 43.5 & 45.8 & \textbf{53.72} & 44.7 & 18.8\\
 & \Ours (GPT-3.5)  & \underline{35.8} & \underline{39.06} & \underline{47.4} & \textbf{52.01} & \textbf{66.8} & \textbf{66.08} & \underline{46.7} & 45.95 & \underline{47.4} & \underline{53.55} & \textbf{51.3} & \textbf{26.3} \\
\bottomrule
\end{tabular}
}
\vspace{-0.15in}
\end{table*}

For simpler queries, we use three benchmark single-hop QA datasets, which consist of queries and their associated documents containing answers, namely \textbf{1) SQuAD v1.1}~\cite{rajpurkar2016squad}, \textbf{2) Natural Questions}~\cite{kwiatkowski2019natural} and \textbf{3) TriviaQA}~\cite{joshi2017triviaqa}.

\paragraph{Multi-hop QA} 
To consider more complex query scenarios, we use two benchmark multi-hop QA datasets, which require sequential reasoning over multiple documents, namely
\textbf{1) 2WikiMultiHopQA (2WikiMQA)}~\cite{ho2020constructing} and \textbf{2) HotpotQA}~\cite{yang2018hotpotqa}. For both single-hop QA and multi-hop QA, we report the accuracy (\textbf{acc}) and F1-score (\textbf{f1}) as evaluation metrics, where acc measures if the predicted answer contains the ground-truth, and f1 measures the number of overlapping words between the predicted answer and the ground-truth.

\paragraph{Long-form QA}
We select an English dataset \textbf{ASQA}~\cite{stelmakh2022asqa}. Specially, we use the ASQA dataset with 948 queries recompiled by ALCE~\cite{gao2023enabling} and apply ALCE's official evaluation metrics, involving String Exact Match (\textbf{str-em}) and String Hit Rate (\textbf{str-hit}). 


\subsection{Baseline\&LLMs}

We extensively compare three types of baselines: 1) No Retrieval (\textbf{NoR}), which directly feeds queries into LLMs to output answers without any retrieval process; 2) Single-time RAG (\textbf{STRAG}), which retrieves knowledge in a one-time setting to answer the original queries; 3) Adaptive RAG (\textbf{ARAG}), which leverages an adaptive forward exploration strategy to retrieve knowledge to enhance answer quality. For STRAG, we select Vanilla RAG, Chain-of-note~\cite{yu2023chain}, Self-Refine, and Self-Rerank are simplified from Self-RAG~\cite{asai2023self}. For ARAG, we include five recent famous methods for comparison - FLARE~\cite{jiang2023active}, Self-RAG, ReAct~\cite{yao2022react}, Adaptive-RAG~\cite{jeong2024adaptive} and Adaptive-Note~\cite{wang2024retriever}. Additionally, we conduct experiments on multiple LLMs, including Qwen2-7b~\cite{Yang2024Qwen2TR}, Llama3-8b~\cite{Touvron2023LLaMAOA} and GPT-3.5 (OpenAI gpt-3.5-turbo-instruct).   We default to using Llama3-8b as the Multi-granular Content Filter LLM, detail experiment setting about multi-filter content see appendix \ref{appendix:filter}. Unless otherwise specified, Llama3-8b was employed as the default model.

\section{Results and Analysis}
In this section, we evaluate our proposed framework, \Ours, on six real-world datasets and compare it against several baselines, including No retrieval, single-time and adaptive RAG methods.
\subsection{Main Results.}
We implemented the \Ours on six datasets. The comparison with baseline models is summarized in Table \ref{tab:2_main_results}. Key observations are as follows:

\paragraph{\Ours vs. Single-time RAG.}
Results show that our method surpassed all STRAG for all six QA datasets. Meanwhile, it is noteworthy that our method outperformed Vanilla by over 30\% on the Natural Questions, 2WikiMQA, and HotpotQA datasets. Even on the remaining three datasets, it still achieved an approximate 10\% improvement. These achievements highlight its superiority and effectiveness. An intuitive explanation is that STRAG heavily depends on the quality of one-time retrieval, whereas our method can adaptively explore more knowledge in the corpus and filter useless chunks and irrelevant sentence in chunk. Therefore, it is able to demonstrate that our method preserves more and more effective knowledge.

\paragraph{\Ours vs. Adaptive RAG.} In Table \ref{tab:2_main_results}, we conduct an in-depth comparison of our approach with several existing ARAG models,including FLARE, Self-RAG, ReAct, Adaptive-RAG and Adaptive-Note. Our method consistently outperforms baselines in single-hop, multi-hop, and long-form QA tasks, particularly in accuracy. Even compared to the state-of-the-art ARAG method, it improves by over 10\%, demonstrating its superiority, effectiveness, and robustness. We provide an in-depth analysis of the baseline limitations and the factors contributing to our success.
\textbf{First}, ReAct and Flare relies on LLMs' internal knowledge to guide retrieval decisions, but its inherent overconfidence~\cite{Zhou2023NavigatingTG} may hinder retrieval efficacy by neglecting existing knowledge.
In contrast, our method employs a greedy strategy to first gather information extensively, followed by a careful assessment of whether to incorporate new, useful knowledge into the existing framework. This process optimizes knowledge extraction and significantly enhances response accuracy.
\textbf{Second}, Self-RAG faces challenges in training effective models for complex tasks due to numerous classifications
such as labeled inputs, retrieved paragraphs, and output categorizations.
Unlike this approach, our Multi-granular Content filter training strategy is relatively simple, yet it maximizes the utilization of valuable information through multiple iterations and agent-based memory.
\textbf{Third}, The Adaptive-RAG method adapts retrieval strategies based on query complexity, and Adaptive-Note generates new memory in each iteration until memory growth stabilizes. However, both of them neglect passage quality, which affects answer accuracy. Instead, our method focuses on the importance of retrieving relevant paragraphs, aiming to minimize the impact of irrelevant information on the LLM's decision-making when answering questions.

\subsection{Classifier Performance}
\begin{figure}[ht]
    \centering
    \hspace{-2cm}
    \begin{subfigure}[t]{0.3\textwidth}
        \includegraphics[width=\textwidth]{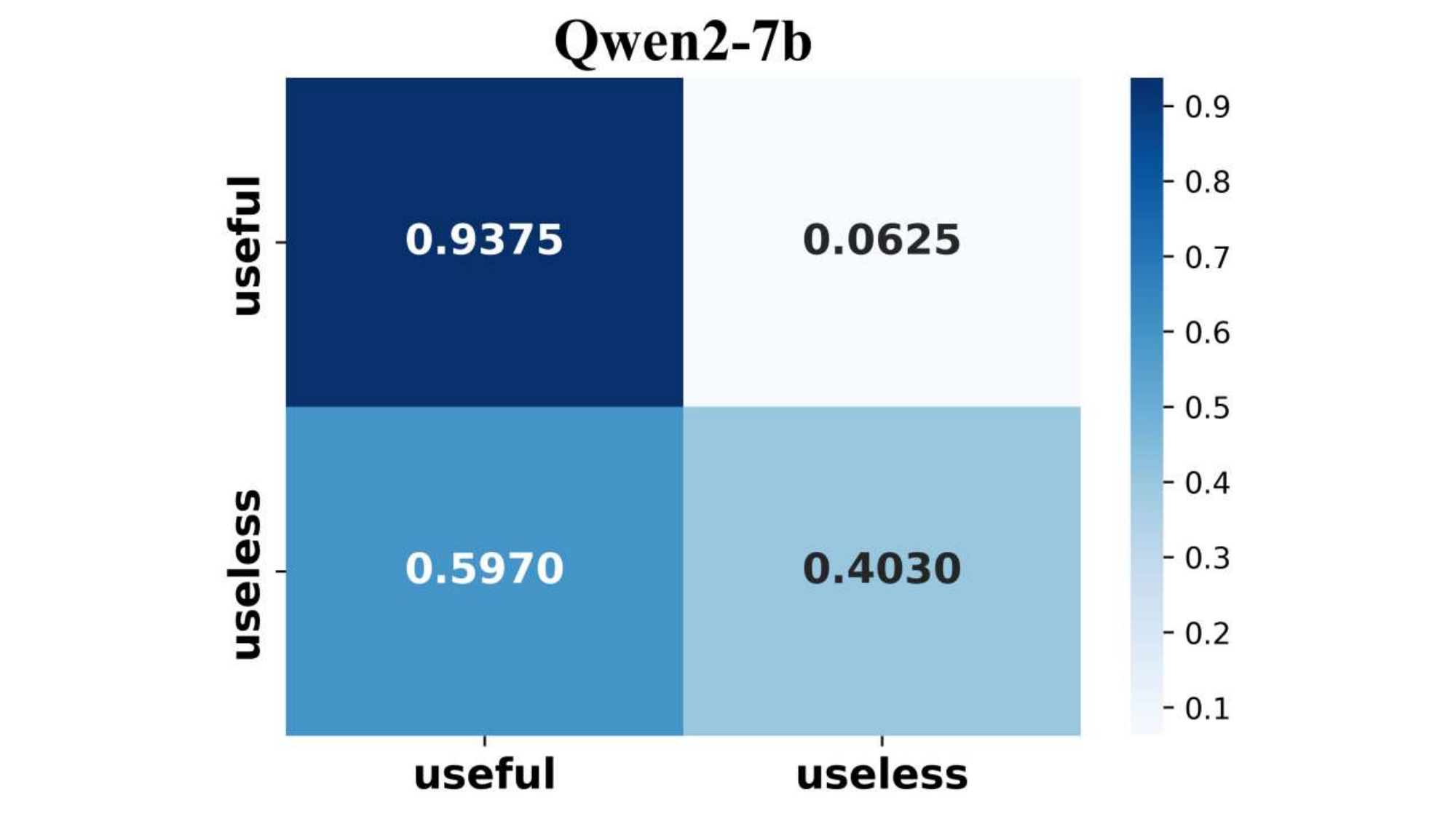}
        \caption{Qwen2-7b}
        \label{fig:4_confusion1}
    \end{subfigure}
    \hspace{-0.8cm}
    \begin{subfigure}[t]{0.3\textwidth}
        \includegraphics[width=\textwidth]{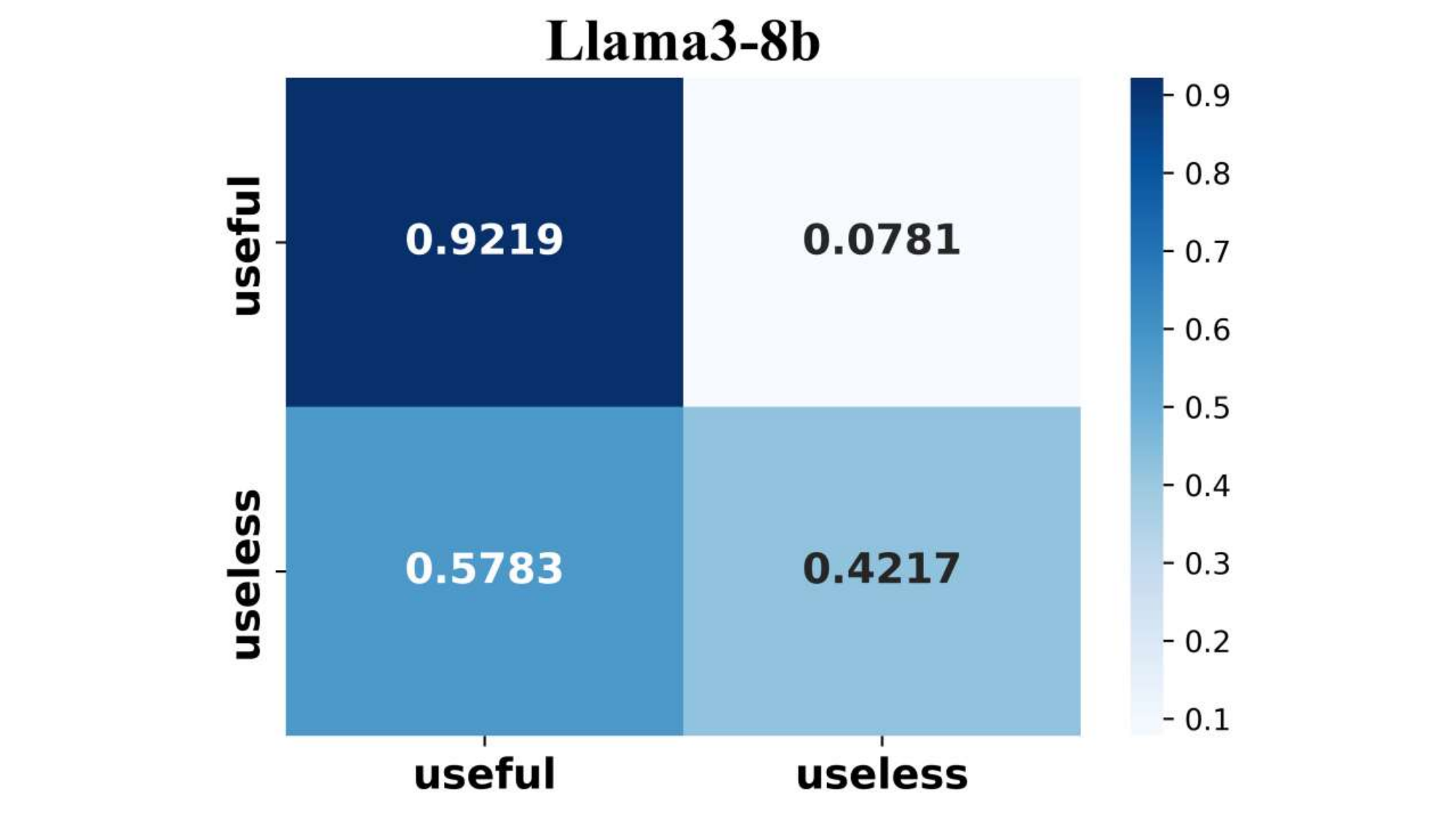}
        \caption{Llama3-8b}
        \label{fig:4_confusion2}
    \end{subfigure}
    \hspace{-2cm}
    \vspace{-0.2cm}
    \caption{\textbf{Confusion matrix for fine-tuned LLMs. Our Fine-Tuned LLMs serve as excellent classifiers.} }
    \vspace{-0.3cm}
    \label{fig:confusion}
\end{figure}

To understand the performance of the proposed classifier, we analyze its effectiveness across two LLM models. As shown in figure \ref{fig:confusion}, whether in Llama3-8b or Qwen2-7b, our \Ours classifier, achieves over 90\% accuracy in classifying useful retrieved passages. Furthermore, it successfully excludes more than 40\% of negative retrieved knowledge, significantly improving the quality of the knowledge and eliminating irrelevant information.
\begin{figure}[ht]
    \centering
    \hspace{-1.2cm}
    \begin{subfigure}[t]{0.23\textwidth}
        \includegraphics[width=\textwidth]{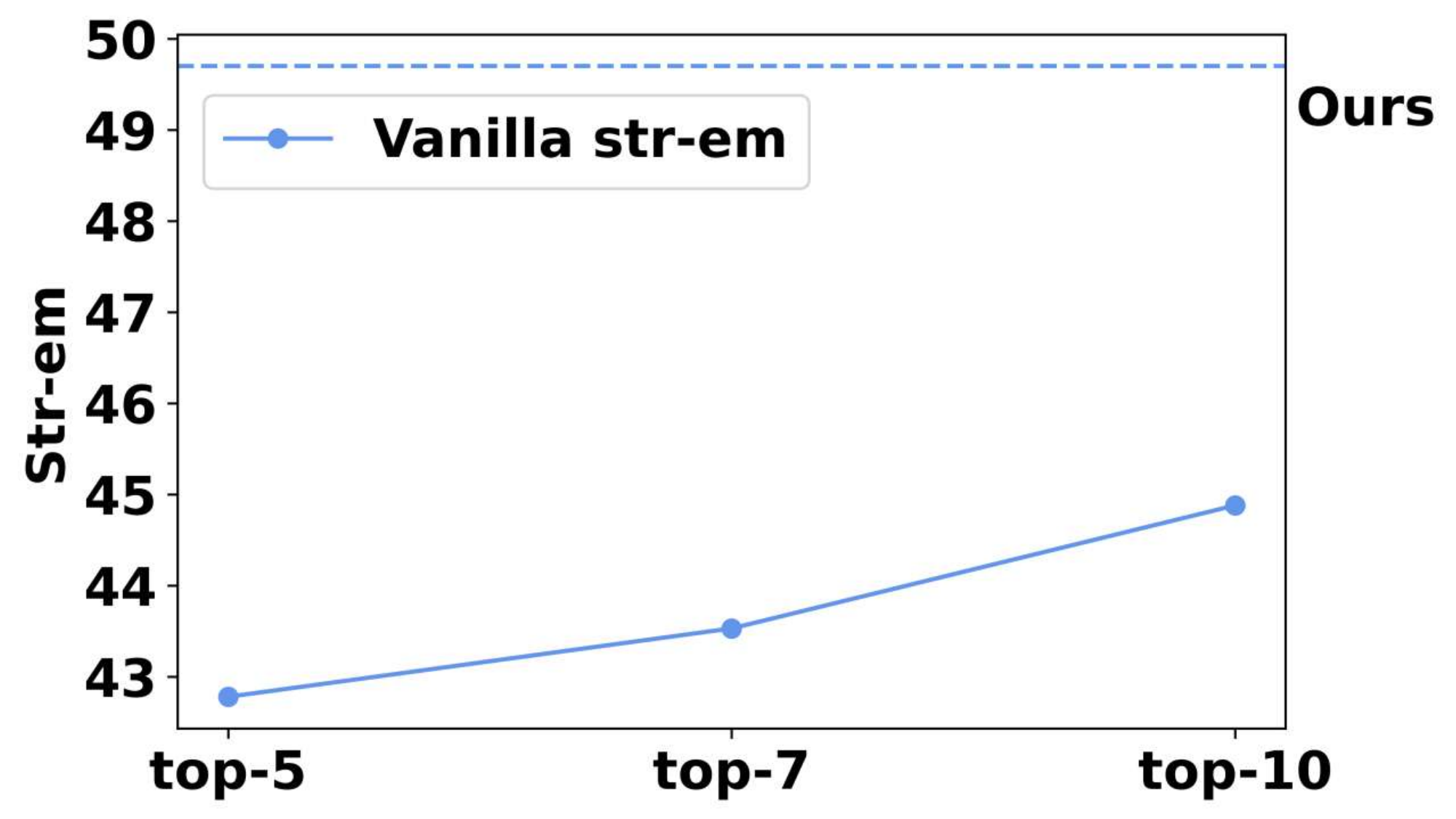}
        \caption{ASQA}
        \label{fig:1_fairtopk}
    \end{subfigure}
    \hspace{0.2cm}
    \begin{subfigure}[t]{0.23\textwidth}
        \includegraphics[width=\textwidth]{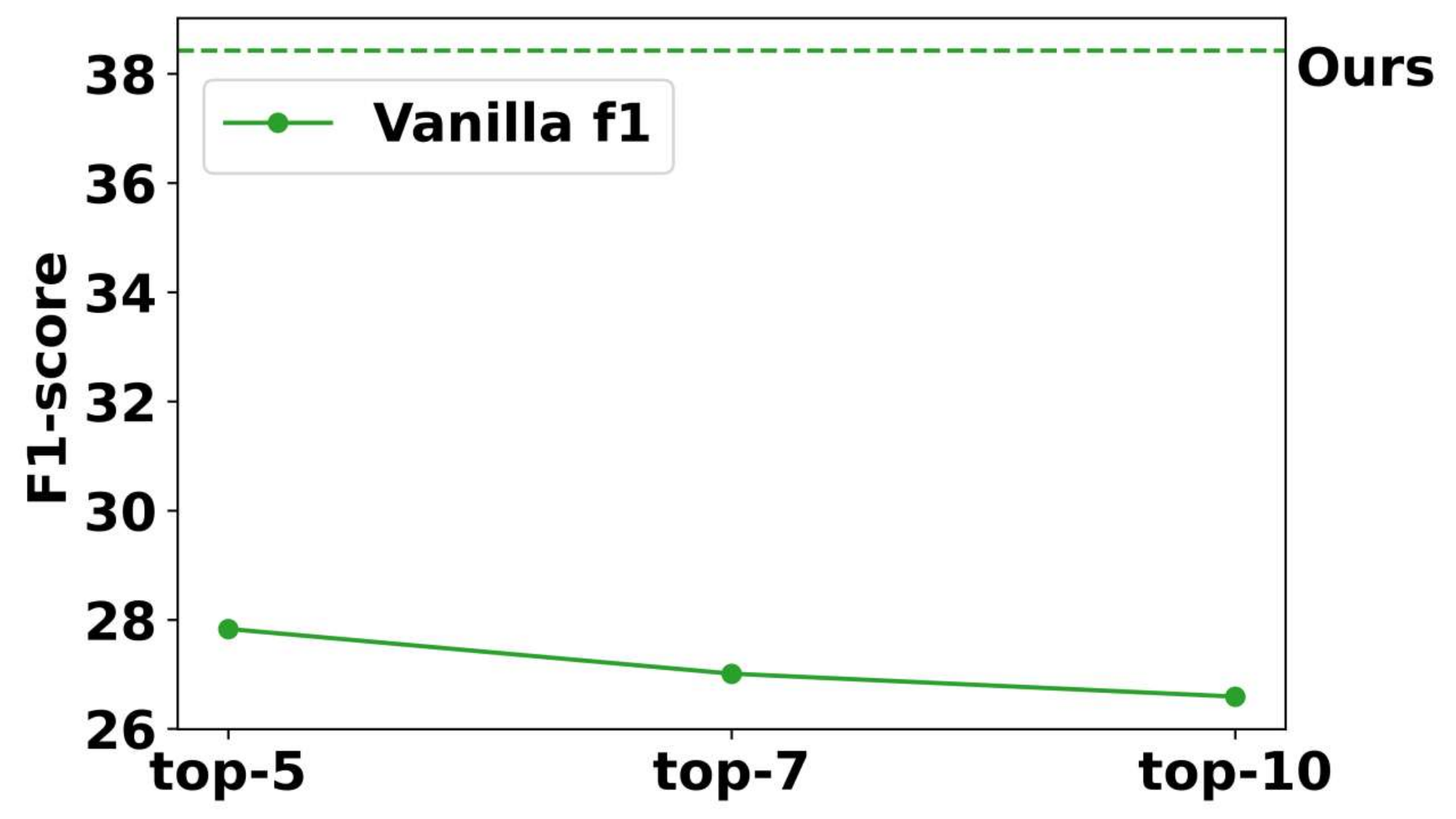}
        \caption{SQuAD}
        \label{fig:2_fairtopk}
    \end{subfigure}
    \hspace{-1.2cm}
    
    \begin{subfigure}[t]{0.23\textwidth}
        \includegraphics[width=\textwidth]{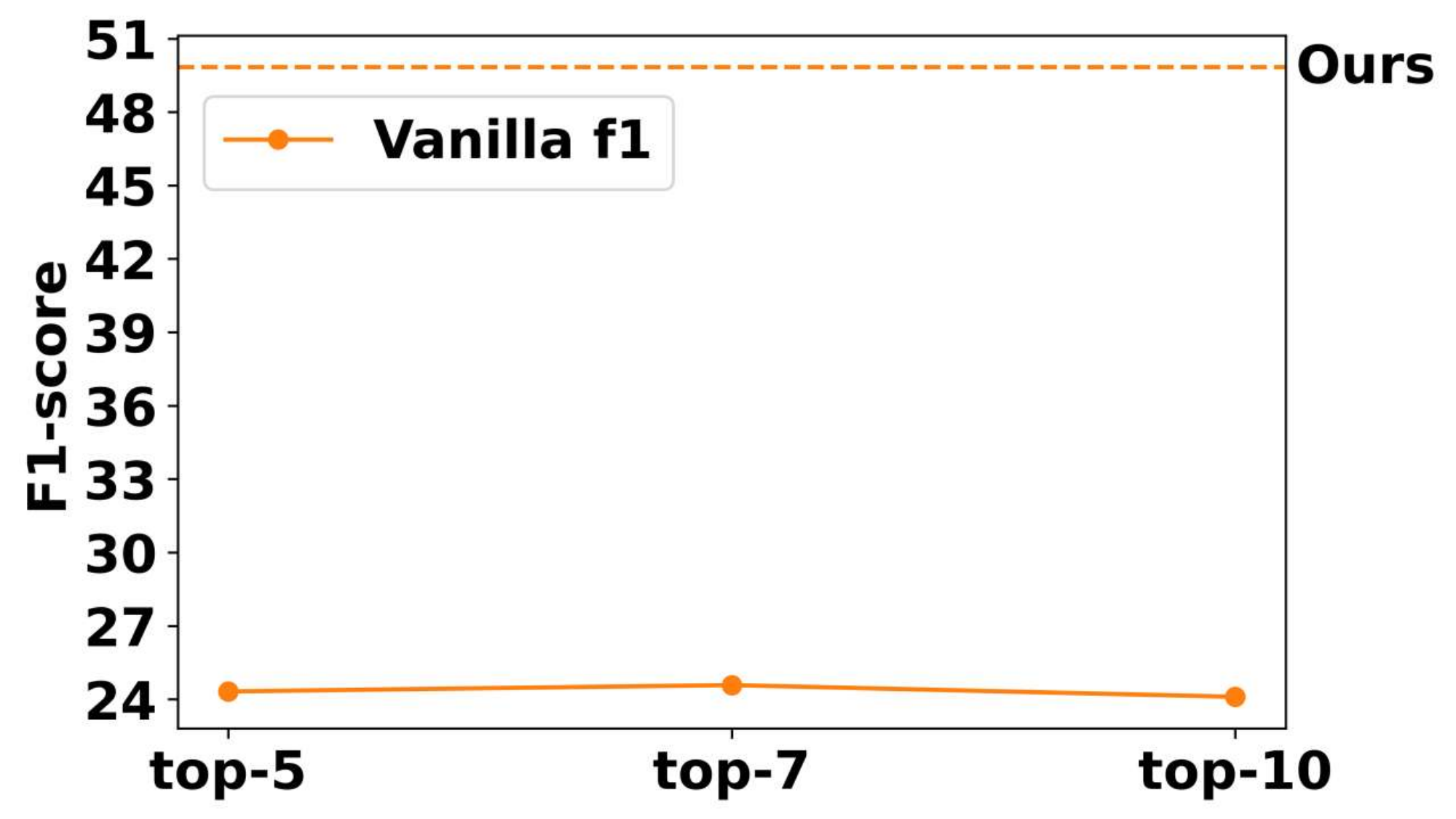}
        \caption{Natural Questions}
        \label{fig:3_fairtopk}
    \end{subfigure}
    \hspace{0.15cm}
    \begin{subfigure}
    [t]{0.23\textwidth}
        \includegraphics[width=\textwidth]{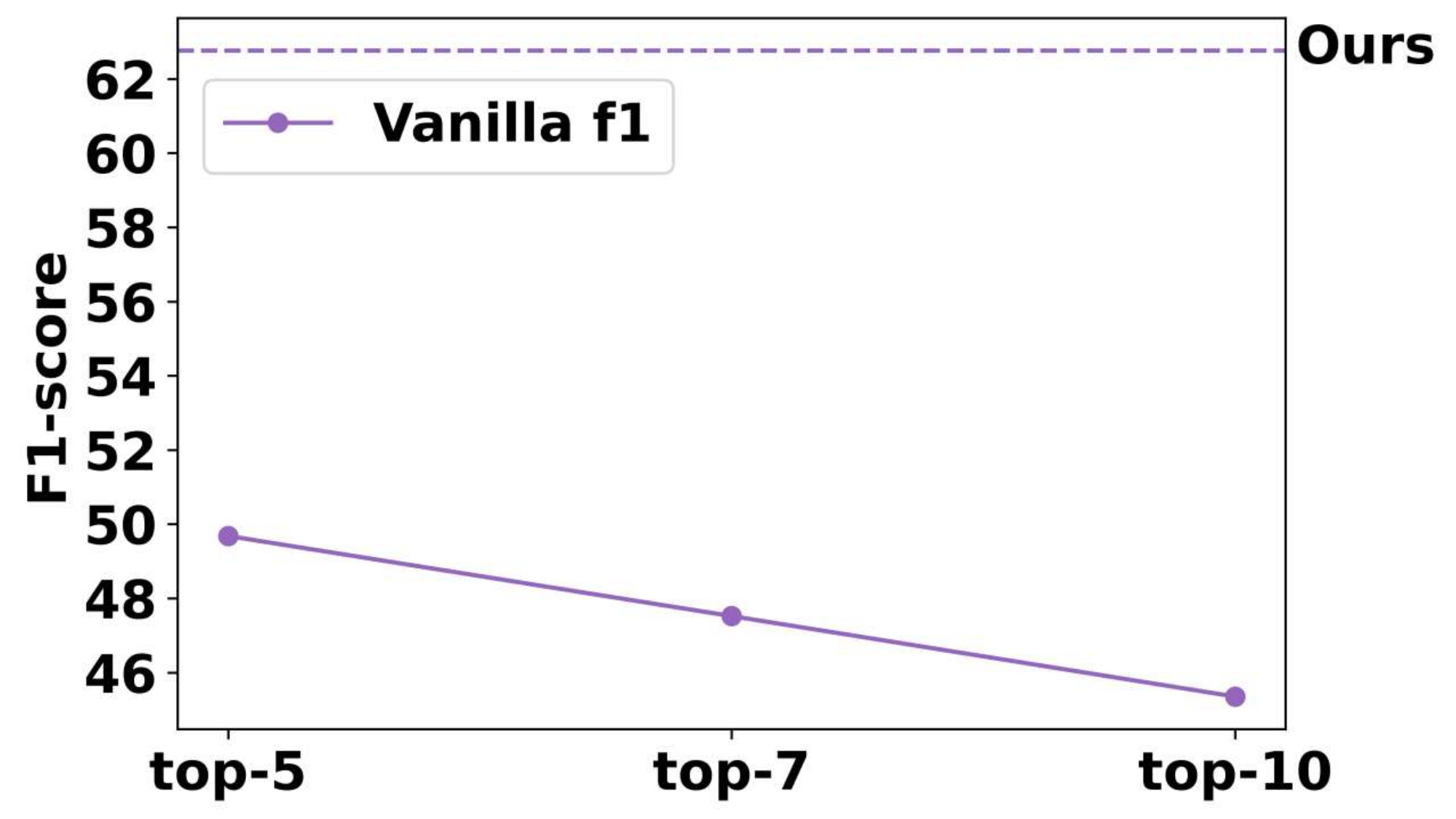}
        \caption{TriviaQA}
        \label{fig:4_fairtopk}
    \end{subfigure}
    \hspace{-1.2cm}
    \begin{subfigure}[t]{0.23\textwidth}
        \includegraphics[width=\textwidth]{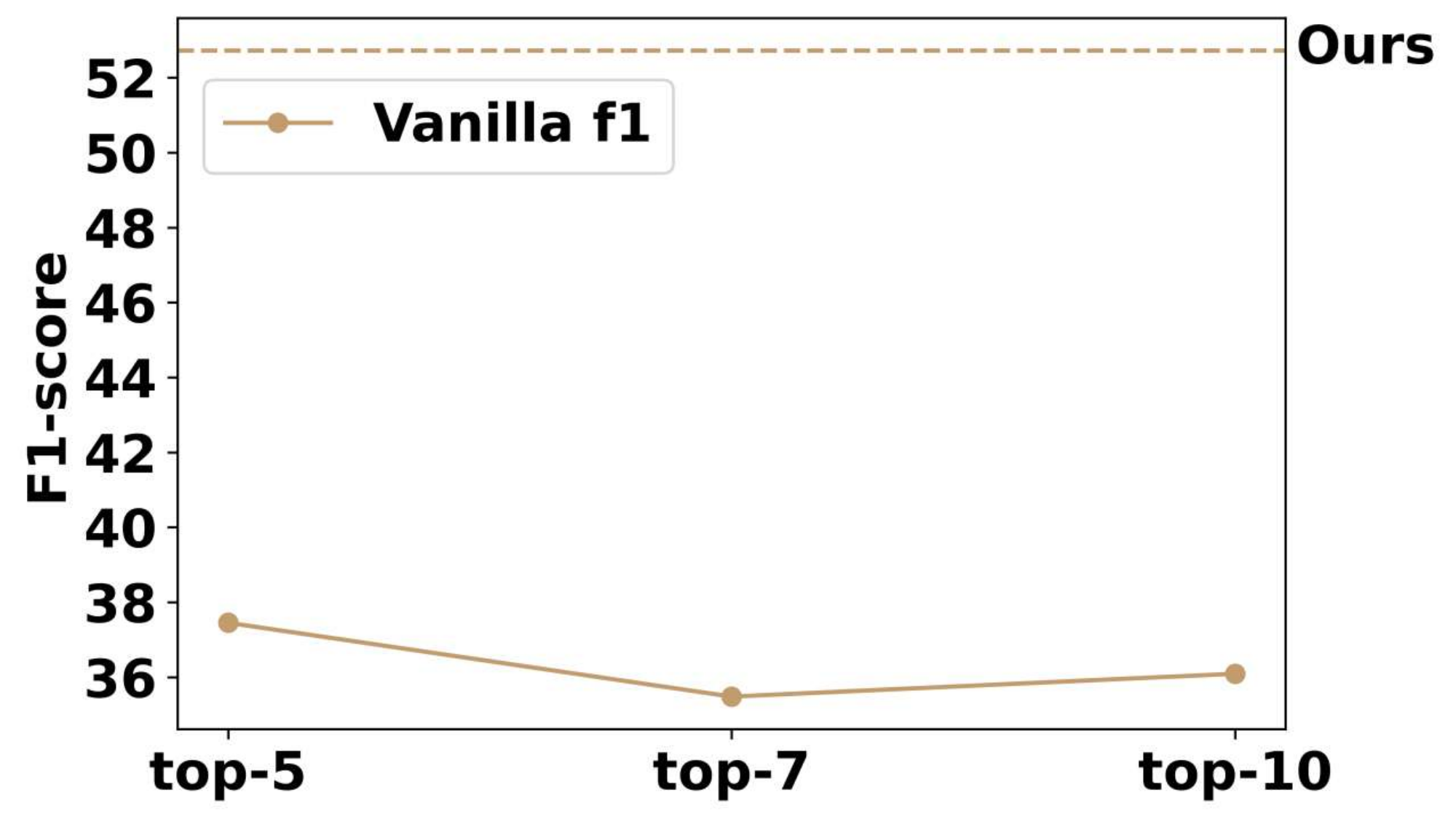}
        \caption{2WikiMQA}
        \label{fig:5_fairtopk}
    \end{subfigure}
    \hspace{0.15cm}
    \begin{subfigure}[t]{0.23\textwidth}
        \includegraphics[width=\textwidth]{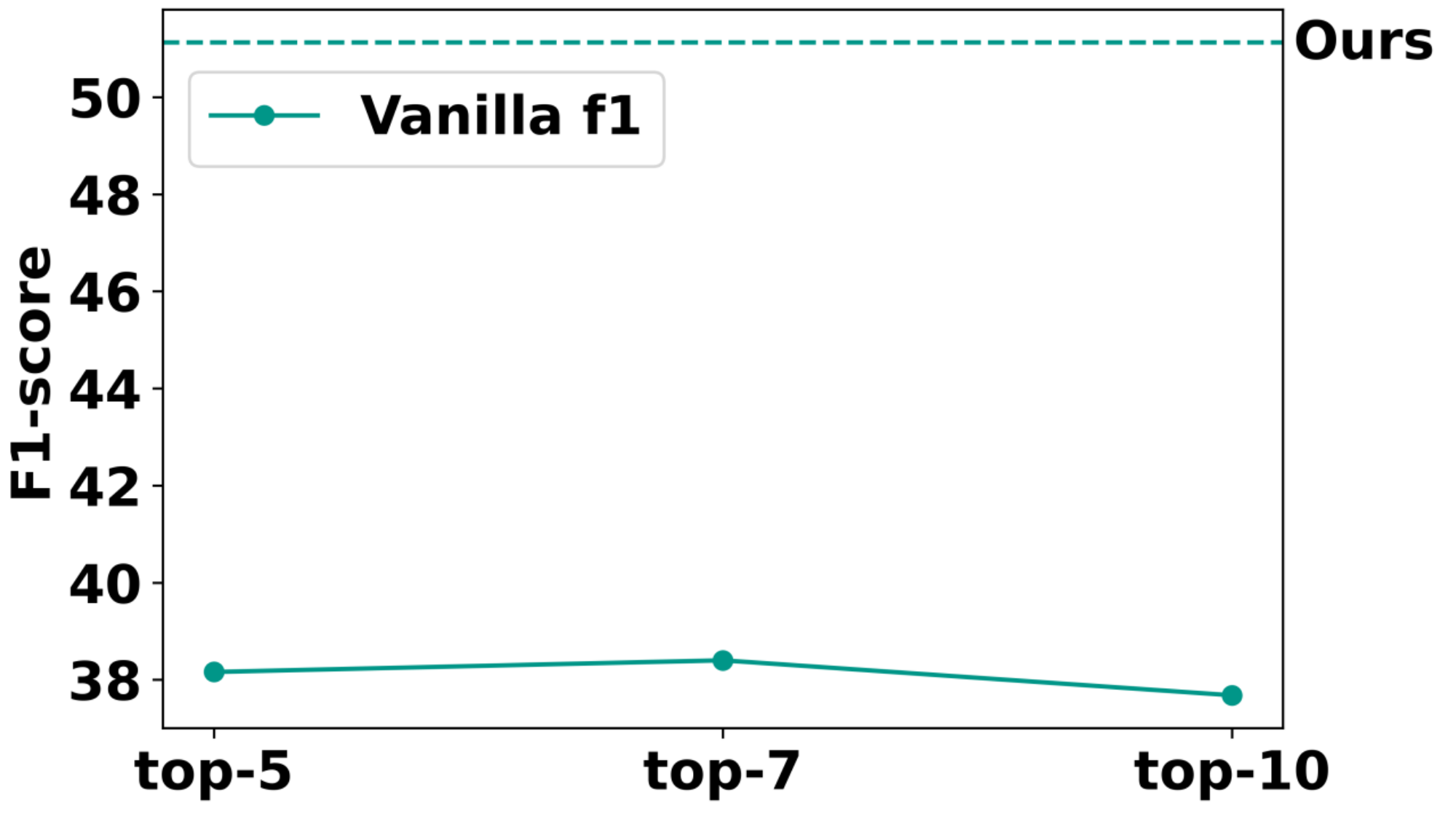}
        \caption{HotpotQA}
        \label{fig:6_fairtopk}
    \end{subfigure}
    \caption{\textbf{Results of the in-depth comparison under a fair \textbf{top-$\boldsymbol{k}$}.}}
    \vspace{-0.7cm}
    \label{fig:Fair_top-k_1}
\end{figure}

\subsection{In-depth comparison under a fair top-k}
Under the same raw top-k setting, ARAG methods generally retrieve more passages compared to single-step methods. Unfortunately, while we can specify the top-k value for each step, the inherent retrieval uncertainly in ARAG prevents us from controlling the total number of retrieved passages. To ensure a fairer performance comparsion, we address the discrepancy by calculating the average number of unique passages retrieved per sample across all adaptive steps, which we term the fair top-k. Figure \ref{fig:Fair_top-k_1} illustrates the overall performance of Vanilla RAG under this fair-top-k setting. It is evident that as the number of retrieved passages increases, the Vanilla method shows little to no significant improvement. These findings further emphasize the superiority of our method.

\begin{table}[t]
  \centering
    \setlength{\tabcolsep}{6pt}
    \fontsize{7 }{9}\selectfont
    \begin{tabular}{ccccccc}
    \toprule
    \textbf{LLM} & \multicolumn{2}{c}{Qwen2-7b} & \multicolumn{2}{c}{LLama3-8b} & \multicolumn{2}{c}{GPT3.5}\\
    \midrule
    metric & acc & f1 & acc & f1 & acc & f1\\
    \midrule
    \Ours & \textbf{56.0} & \textbf{52.73} & \textbf{43.8} & \textbf{43.5} & \textbf{46.7} & \textbf{45.95}\\
    \midrule
    \rowcolor{gray!20}
    \multicolumn{7}{c}
    {\textbf{Multi-granular Content Filter}} 
    \vspace{0.1cm}
    \\
    w/o CF & 52.2  & 49.78  & 41.5  & 40.51 & 43.5  & 42.37 \\
    w/o SF & 55.1  & 50.67 & 42.8  & 42.17  & 45.2  & 43.84 \\
    w/o ALL & 51.6 & 48.39 & 40.9 & 40.12 & 42.4 & 41.65 \\
    \midrule
    \rowcolor{gray!20}
    \multicolumn{7}{c}{\textbf{Agent-Based Memory  Updater}} \vspace{0.1cm}\\
    w/o AM & 53.5  & 50.97  & 42.7  & 41.63 & 44.3  & 42.85 \\
    \bottomrule
    \end{tabular}%
    \vspace{-0.2cm}
      \caption{\textbf{Ablation study on the 2WikiMQA dataset.}}
      \vspace{-0.4cm}
  \label{tab:ablation}%
\end{table}%

\subsection{Ablation Study}

To analyze the contributions of components in the proposed \Ours method, particularly the fine-tuning of the Multi-granular Content Filter and the Agent-based Memory Updater in Adaptive information Collector, we conducted an ablation study on the 2WikiMQA dataset. The above components comprise three key components:
\begin{itemize}[left=0em, itemsep=-5pt, topsep=5pt]
    \item \textbf{Chunk-Level Filter (CF):} Input: Query and retrieved chunk. Output: useful / useless.
    \item \textbf{Sentence-Level Filter (SF):} Input: query and filtered chunk. Output: filtered sentences.
    \item \textbf{Agent-Based Memory Updater (AMU): } whether to use agent-based method for memory.
\end{itemize}

\paragraph{Effect of the Multi-granular Content Filter.}  
To evaluate the contribution of each Multi-granular Content Filter component, we systematically removed one type of content. Additionally, removing both two types (\textit{w/o ALL}) effectively disables the content filter stage. The results, presented in Table \ref{tab:ablation}, show that the model achieves its best performance when both two filter strategy are used together. Conversely, removing any single type of data leads to a noticeable decline in performance, highlighting the importance of each component in enhancing the framework’s overall effectiveness. Interestingly, the performance when Sentence-Level Filter (\textit{w/o SF}) are excluded remains higher than when Chunk-Level Filter (\textit{w/o CF}) are omitted. This indicates that, The Chunk-Level Filter plays a more significant role in our approach, effectively filtering out irrelevant chunk information to a large extent.
\paragraph{Impact of Agent-based Memory  Updater.}  
To examine the role of Agent-based Memory Updater(\textit{w/o AMU}) in \Ours, we conducted an ablation experiment by removing the AMU module. As shown in Table \ref{tab:ablation}, removing this module significantly decreases performance. This highlights the agent’s critical role in memory generation. The agent ensures the creation of more efficient memory, thereby enabling the LLM to provide more accurate responses.

The ablation study highlights the importance of each content filter component and the agent-based memory module. Multi-granular Content Filter and Agent-based Memory Updater significantly enhance the performance of the \Ours framework.

\begin{table}[t]
  \centering
    \setlength{\tabcolsep}{6pt}
    \fontsize{7 }{9}\selectfont
    \begin{tabular}{ccccccc}
    \toprule
    \multirow{2}[1]{*}{\textbf{top-$\boldsymbol{k}$}} & \multicolumn{2}{c}{2WikiMQA} & \multicolumn{2}{c}{HotpotQA} & \multicolumn{2}{c}{ASQA}\\
       & (acc) & (f1) & (acc) & (f1) & (str-em) & (str-hit) \\
    \midrule
    \rowcolor{gray!20}
    \multicolumn{7}{c}
    {\textbf{Vanilla}} \\
    top-3 & 36.3  & 36.82  & 35.9  & 37.8 & 42.5  & 17.5 \\
    top-5 & \textbf{37.0}  & \textbf{37.45} & 37.6  & 38.16  & 42.78  & 18.14 \\
    top-7 & 35.6 & 35.48 & \textbf{39.8} & \textbf{38.4} & \textbf{43.53} & \textbf{17.93} \\
    \midrule
    \rowcolor{gray!20}
    \multicolumn{7}{c}{\textbf{Ours}} \\
    top-3 & 53.2  & 40.3  & 50.9  & 49.4 & 48.3  & 23.5 \\
    top-5 & \textbf{56.0} & \textbf{42.73} & 52.6  & \textbf{51.13} & 49.7  & 25.2  \\
    top-7 & 55.2  & 41.7  & \textbf{52.8} & 51.02 & \textbf{50.1} & \textbf{25.6} \\
    \bottomrule
    \end{tabular}%
    \vspace{-0.2cm}
      \caption{\textbf{\Ours with different top-$\boldsymbol{k}$ with Qwen2-7b}.}
  \label{tab:Different Top-k}%
\end{table}%
\begin{table}[t]
  \centering
    \setlength{\tabcolsep}{6pt}
    \fontsize{7}{9}\selectfont
    \begin{tabular}{ccccccc}
    \toprule
    \multirow{2}[2]{*}{\textbf{max-iter}} & \multicolumn{2}{c}{2WikiMQA} & \multicolumn{2}{c}{HotpotQA} & \multicolumn{2}{c}{ASQA} \\
     & (acc) & (f1) & (acc) & (f1) & (str-em) & (str-hit) \\
    \midrule
    1 & 53.2 & 38.95 & 50.4 & 49.73 & 45.3 & 20.9 \\
    3 & \textbf{56.0} & \textbf{42.73} & \textbf{52.6} & \textbf{51.13} & \textbf{49.7} & \textbf{25.2} \\
    5 & 55.1  & 41.05 &	51.9 & 50.34 & 48.4 & 24.5 \\

    \bottomrule
    \end{tabular}%
    \vspace{-0.2cm}
      \caption{\textbf{\Ours with varying max iterations in AIC.}}
      \vspace{-0.4cm}
  \label{tab:Adaptive parameter}%
\end{table}%

\subsection{Parameter Analysis}
\paragraph{Impact of top-k.} In Table \ref{tab:Different Top-k}, we present a comparison between our method and Vanilla RAG across different top-k settings. The results demonstrate that our approach consistently outperforms Vanilla RAG under the same top-k conditions, highlighting its robustness in achieving reliable improvements regardless of the number of retrieved passages. Additionally, in most cases, the performance of our system improves as the number of retrieved passages (top-k) per step increases.

\paragraph{Impacts of max iterations} 
As shown In Table \ref{tab:Adaptive parameter}, performance on these complex QA datasets improves with increasing iterations, peaking at 3. We therefore recommend setting the max iterations to 3 for optimal performance.

\section{Conclusion}
In this paper, we propose a novel RAG method, \Ours, based on memory-adaptive updates. Our approach introduces a collaborative multi-agent memory updating mechanism, combined with an adaptive retrieval feedback iteration and a multi-granular filtering strategy. This design enables efficient information gathering and adaptive updates, significantly improving answer accuracy while reducing hallucinations. We validated \Ours and its core components across several open-domain QA datasets. Extensive experiments
prove the superiority and effectiveness of \Ours  .

\paragraph{Limitation.}
Although \Ours has made significant progress in open-domain question answering with RAG, there are still some limitations. First, the framework requires multiple fine-tuning steps to train the Multi-granular Content Filter, which necessitates the collection of a substantial amount of data, as well as considerable computational resources and time. Second, since \Ours requires multiple accesses to the LLM, answering each question takes more time compared to the vanilla approach. In future work, we plan to design a more time-efficient and generalizable fine-tuning strategy to improve the quality of open-domain question answering, thereby enhancing the overall effectiveness of \Ours.

\bigskip

\cleardoublepage

\appendix
\section{Multi-granular Content Filter}\label{appendix:filter}
In \Ours, for the Multi-granular content filter, we finetune the LLaMA 3-8B and Qwen 2-7B models using the LLamaFactory framework. Specifically, for the query-reference classification task, we finetune the model into two categories: useful and useless, retaining only the useful ones. For the context filter, we use only the extracted content from the original passages and feed it into the LLMs. Both fine-tuning processes are conducted over 2 epochs, with the per-device training batch size set to 4.

\paragraph{Chunk-Level Filter}
The accuracy  of responses generated by LLMs can be significantly compromised by noisy retrieved contexts~\cite{yoran2023making}. To mitigate this, we introduce the Chunk-Level Filter module to enhance response accuracy and robustness. This module utilizes LLMs to filter out irrelevant knowledge. Rather than directly querying an LLM to identify noise, we incoporate a Natural Language Inference (NLI) framework~\cite{bowman2015large} for this purpose. Specially, for a query  \(q\) and retrieved reference \(r\), the NLI task evaluates whether the knowledge contains reliable answers, or usefule information aiding the response to the question. This results in a judgment \(j\) categorized as useful or useless. The operation of the Chunk-Level Filter can be mathematically represented as :
\begin{equation}
    F_{\theta}(q, r) \rightarrow j \in \{\text{useful}, \text{useless}\}
\end{equation}
Knowledge is retained if the NLI result is classified as useful, and the reference is discarded when the NLI result is classified as useless. The NLI training dataset is constructed semi-automatically. We provide task instruction, query \(q\), along with retrieved reference \(r\) as prompt to GPT-4, which then generated a brief explanation \(e\) and a classification result \(j\).
The prompt template is as follows:

\begin{tcolorbox}
    \textbf{[Instruction]:} Your task is to solve the NLI problem: given the
premise in [Knowledge] and the hypothesis that "The [Knowledge]
contains reliable answers aiding the response to [Question]". You
should classify the response as useful and useless.

    \textbf{[Question]:}\{Question\}
    
    \textbf{[Knowledge]:}\{Knowledge\}
    
    \textbf{[Format]:}\{Explanation\}\{NLI result\}
\end{tcolorbox}

\paragraph{Sentence-Level Filter}
Followed by previous work~\cite{wang2023learning}, we use the STRINC measure for single-hop QA datasets and CXMI for multi-hop datasets. We train the sentence-Level Filter models \(M_{slf}\), using context filtered with above two measures. To create training data for \(M_{slf}\), for each training sample with query \(q\), we concatenate the retrieved passages \(P\) and query \(q\), then, we apply the filter method \(f\) to obtain filtered context \(t_{output}\) as output. The \(s\) is the sentence of the retrieved passages, and \(t_{output}\) can be represented as:
\begin{equation}
    t_{\text{output}} = \left[ s_{text} \mid s_{strinc} == 1 \right] 
\end{equation}
\begin{equation}
    t_{\text{output}} = \left[ s_{text} \mid s_{cxmi} >= threshold \right] 
\end{equation}

We train \(M_{slf}\) by feeding in query and retrieved passages \(P\), and ask it to generate filtered context.
\section{Retriever \& Corpus}\label{appendix:retriever}
To ensure a fair comparison of all baselines, we align the retriever and corpus across all methods for each dataset. For both single-hop and multi-hop datasets, we employ BM25~\cite{robertson1995okapi}, implemented in the search tool Elasticsearch, as the foundational retriever. For the external document corpus, we use the Wikipedia corpus preprocessed by ~\cite{karpukhin2020dense} for single-hop datasets, and the preprocessed corpus by ~\cite{trivedi2022interleaving} for multiple-hop datasets. For long-form ASQA dataset, we employ dense retriever GTR-XXL~\cite{ni2021large} and use the corpus provided by ALCE, consisting of the 2018-12-20 Wikipedia snapshot, segmented into 100-word passages.
\section{Implementation Details}
For computing resources, we utilize NVIDIA 4090 GPUs with 24GB of memory. Additionally, due to the frequent access to the LLM, we employ VLLM as the inference framework. The software stack includes Python 3.10.15, VLLM 0.6.3.post1, PyTorch 2.5.0, and CUDA 12.1.
\section{Detailed prompt}



We present all the prompts used in our method in Tables A3 and A4. In Table A3, we detail the prompt for the Multi-granular Content Filter. Specifically, at the Memory Initialization stage, {query} represents the original query 
q, and {refs} refers to the retrieved 
k passages obtained by feeding the original query q into the retriever. At the Iterative Information Collection stage, {query} still represents the original query q, and {note} refers to the content of the optimal memory $M_{opt}$. Additionally, as mentioned in the main text, LLMs tend to ask similar questions if previous ones were not well resolved. To address this, we introduce the already-asked questions list {query log} to avoid repetition. At the Note-Updating stage, {query} still refers to q, while {refs} represents new retrieved k passages based on the updated queries, and {note} refers to $M_{opt}$. In the Memory Updating phase, {query} represents the original query q, while {best note} and {new note} represent 
$M_{opt}$ and $M_{cur}$, respectively. 
\\
In the Multi-granular Content Filter stage, for the Chunk-Level Filter, {External\_knowledge} refers to the retrieved k passages, from which we filter out useless passages, retaining only the useful ones. Next, for the Sentence-Level Content Filter, {context} refers to each useful passage. After passing through this filter, we extract important sentences from the passages to generate the answer.
\\
In the Agent-based Memory Update, we assume three roles in the memory generation process: reviewer, challenger, and refiner. The reviewer evaluates the strengths and weaknesses of the note memory based on the query and {refs}. The challenger, using the reviewer’s feedback, provides suggestions to revise and enhance the memory. Finally, the refiner uses both the reviewer’s insights and the challenger’s suggestions to refine and generate the new memory. In the final memory updating phase, we compare the new memory with the initialized memory to select the best memory.
\begin{table*}[h]
  \centering
  \begin{tabular}{|p{16cm}|}
    \hline
    \multicolumn{1}{|>{\columncolor{gray!10}}c|}
    {\textbf{Prompt of the Memory Initialization Stage}} 
    \\
    \hline
    \textbf{Instruction:} \\
    Based on the provided document content, write a note. The note should integrate all relevant information from the original text that can help answer the specified question and form a coherent paragraph. Please ensure that the note includes all original text information useful for answering the question. \\
    \\
    Question to be answered: \{query\} \\
    Document content: \{refs\} \\
    \\
    Please provide the note you wrote: \\
    \hline
    \multicolumn{1}{|>{\columncolor{gray!10}}c|}{\textbf{Prompt of the Iterative query rewritten Stage}} \\
    \hline
    \textbf{Instruction:} \\
    Task: Based on the notes, propose a new question. The new question will be used to retrieve documents to supplement the notes and help answer the original question. The new question should be concise and include keywords that facilitate retrieval. The new question should avoid duplication with the existing question list. \\
    \\
    Original question: \{query\} \\
    Notes: \{note\} \\
    Existing question list: \{query\_log\} \\
    \\
    Provide your new question,you MUST reply with the new question on the last line, starting with "\#\#\# New Question". \\
    \hline
    \multicolumn{1}{|>{\columncolor{gray!10}}c|}{\textbf{Prompt of the Chunk-Level Filter}} \\
    \hline
    \textbf{Instruction:} \\
  You are an advanced AI model specialized in understanding the Natural Language Inference (NLI) tasks. Your task is to do the NLI problem. The premise is [External Knowledge]. The hypothesis is "There exist clear and unambiguous answer in the [External Knowledge] that can convincingly and soundly answer the Question." Your response should be in one of (useful,useless). \\
    \\
    External Knowledge: \{External\_Knowledge\} \\
    Question: \{Question\} \\
    \\
    Now give me the NLI result, which 1. should be one of (useful,useless). 2.Please strictly following this json format and fill xxx with your answer. 3. Please notice the Escape Character and keep correct format. 4. Please just give me the concise Json response and no ther redundant words. 5. The output should not appear Here is the NLI result, Just strictly follow the format below: \\
    \{"NLI result":"xxx"\}
        \\
    \hline
    \multicolumn{1}{|>{\columncolor{gray!10}}c|}{\textbf{Prompt of the Sentence-Level Filter}} \\
    \hline
    \textbf{Instruction:} \\
   You are an AI model specialized in extracting helpful sentences from a given context. Your task is to extract helpful sentences while removing irrelevant or unhelpful ones based on the provided question and context.
   
    \\
    Question: \{query\} \\
    context: \{context\} \\
    \\
    Now provide the extracted helpful sentences, which should include only valid and relevant sentences from the context. \\
    \hline
  \end{tabular}
  \caption{All prompts of Memory initialize, query rewritten and Multi-granular Content Filter.}
  \label{tab:All prompts of IIC and AMR}
\end{table*}

\begin{table*}[h]
  \centering
  \begin{tabular}{|p{16cm}|}
    \hline
    \multicolumn{1}{|>{\columncolor{gray!10}}c|}
    {\textbf{Prompt of the reviewer in Agent-based memory}} 
    \\
    \hline
    \textbf{Instruction:} \\
    Task: Analyze the relationship between the query, retrieved documents, and notes. Identify the strengths and weaknesses of how well the notes align with the query and incorporate the information from the retrieved documents. Highlight areas where the notes effectively cover the query and the references, as well as areas where they could be improved to better address the query or utilize the information from the references.\\
    Question: \{query\} \\
    retrieved documents: \{refs\} \\
    note: \{note\}
    \\
    Provide an analysis of the notes with a focus on the strengths and weakness: \\
    \hline
    \multicolumn{1}{|>{\columncolor{gray!10}}c|}{\textbf{Prompt of the challenger in Agent-based memory}} \\
    \hline
    \textbf{Instruction:} \\
    Based on the provided reviewer information, provide specific and actionable suggestions to improve the notes. The goal is to ensure the notes comprehensively and accurately address the query while fully utilizing relevant information from the retrieved documents. \\
    Question: \{query\} \\
    retrieved documents: \{refs\}\\
    Notes: \{note\} \\
    reviewer information: \{review\_info\} \\
    
    Provide detailed suggestions to revise and enhance the notes: \\
    \hline
     \multicolumn{1}{|>{\columncolor{gray!10}}c|}{\textbf{Prompt of the refiner in Agent-based memory}} \\
    \hline
    \textbf{Instruction:} \\
    Refine the provided notes based on the reviewer information and suggestions. The goal is to ensure the notes are improved to better address the query and fully utilize the relevant information from the retrieved documents. \\
    Question: \{query\} \\
    retrieved documents: \{refs\}\\
    Notes: \{note\} \\
    reviewer information: \{review\_info\} \\
    suggestions: {suggestions}
    \\
    Provide the refined notes that incorporate the feedback from the reviewer information and suggestions: \\
    \hline
    \multicolumn{1}{|>{\cellcolor{gray!10}}c|}{\textbf{Prompt of the memory updating}} \\
    \\
    \textbf{Instruction:}
    \newline{}
    Task: Please help me determine which note is better based on the following evaluation criteria:
    \newline{}
    1. Contains key information directly related to the question.
    \newline{}
    2. Completeness of Information: Does it cover all relevant aspects and details?
    \newline{}
    3. Level of Detail: Does it provide enough detail to understand the issue in depth?
    \newline{}
    4. Practicality: Does the note offer practical help and solutions?
    \newline{}
    Please make your judgment adhering strictly to the following rules:
    \newline{}
    - If Note 2 has significant improvements over Note 1 based on the above criteria, return 
    \{``status": ``True"\}
    directly; otherwise, return
    \{``status": ``False"\}
    .
    \newline{}
    Question: \{query\}
    \newline{}
    Provided Note 1: \{best\_note\}
    \newline{}
    Provided Note 2: \{new\_note\}
    \newline{}
    Based on the above information, make your judgment without explanation and return the result directly.
    \\
    \hline
  \end{tabular}
  \caption{All prompts of Agent-based Memory Update}
  \label{tab:All prompts of Agent-based Memory Update}
\end{table*}

\end{document}